%% file: lettre_Nature_0198.tex
\documentstyle[aps,12pt,preprint,epsfig]{revtex}

\begin{document}
\def\bra{\langle}
\def\ket{\rangle}
\def\ie{{\it i.e.}, }
\def\pref#1{(\ref{#1})}

\title{\bf Moisture induced Ageing in Granular Media}

\author{L. Bocquet, E. Charlaix$^{(\dag)}$, S. Ciliberto, J. Crassous}

\address{Laboratoire de Physique de l'E.N.S. de Lyon (URA CNRS 1325), \\
46 All\'ee d'Italie, 
69364 Lyon Cedex, France \\
and $~^{(\dag)}$ DPM (UMR CNRS 5586), Universit\'e 
Claude Bernard-Lyon I, \\43 Bd du 11 Novembre 1918, 69622 Villeurbanne Cedex, France.}

\maketitle

{\bf We present experiments showing that a granular system of small 
beads does exhibit ageing properties : its maximum stability angle is
measured to increase logarithmically with resting time, \ie the time
elapsed before performing the measure. We show that humidity is the crucial 
ingredient responsible for this behaviour : while ageing effects are 
important at intermediate humidity, they 
disappear at vanishing humidity. On the basis of these experimental results, we propose a model based on the activated
condensation of liquid bridges between the beads. Within this picture, 
we are able to reproduce both the waiting-time and humidity dependence of the 
ageing properties.}

In his pioneering treatise published in 1773 \cite{Coulomb},
Charles de Coulomb pointed out the peculiar properties of granular
systems. 
His decisive remark was to recognize the links existing between the
statics of granular systems and the friction properties of two solids
at contact : the equilibrium and stability of granular edifices could
be discussed in terms of the friction between the different part or
layers of the system \cite{Duran}. In particular, he  deduced a relationship
between the rest angle of the pile, $\theta_0$, and the static friction 
coefficient $\mu_s$ of the medium, in the form $\tan \theta_0=\mu_s$.
One knows by now that this relation is only
a good first approximation and that a wide variety of behaviours can be
observed in granular systems \cite{Duran}. The same
is true for dry friction problems : while the subject remained  
poorly understood during more than 150 years, a lot of new
results emerged in the last four decades. In particular, one of the intriguing
behaviour of friction is
the existence of ageing properties of the static friction coefficient, $\mu_s$ :
the latter is found to increase logarithmically with resting time,
\ie with the time during which the system was left at rest before 
pulling it (in order {\it e.g.} to probe $\mu_s$).
Now, if Coulomb's analogy between solid
 friction and the statics of granular systems is further pursued,  a 
 question arises : can ageing be observed in granular systems too ?

In order to verify this assertion, we have studied the effect of waiting
time on the angle of first avalanche $\theta_w$ of a granular system of 
small (typically $200 \mu m$) spherical glass beads contained in a rotating drum, following a protocole 
described in fig. 1.a.
Astonishingly, and in close analogy with dry friction, a logarithmic
ageing of 
the maximum static angle $\theta_w$ is observed, as shown in fig. 2.a.
In other words, granular media exhibit ageing. 
The dependence on bead size has been studied and ageing was not
observed for beads with a diameter larger than $0.5~mm$, except
for very large humidities.
On the other hand it is known that humidity is an important
parameter in the description of granular media \cite{Duran}. 
Addition of small quantities of wetting liquid has been shown to 
change enormously the repose angle of a pile \cite{Castle} ; a
 brief discussion of moisture effects can even be found in 
Coulomb's treatise \cite{Coulomb}. We thus repeated the experiments for
various humidities $P_v/P_{sat}$. As shown on figure 2.b,  
humidity appears to be the {\bf crucial parameter} which controls the ageing of 
$\theta_w$ :
no ageing is observed for small humidity, while the ``strength'' $\alpha$ of the
ageing effect increases in an enormous way with humidity. 

This humidity dependence leads to the intuitive idea
that the ageing effects originate from the condensation of 
small liquid bridges between 
the beads.  Indeed, liquid bridges induce a significant
cohesion force between the beads, which can increase the friction
between different layers of the granular system and result in a higher
value of $\theta_w$. 
However,  the physical justification of this
intuitive behaviour is not obvious at all. 
Let us first consider the idealized situation of two smooth beads
in contact : the capillary adhesion force exerted by a small liquid
bridge connecting the beads is given by the product of the liquid
surface tension times the radius of the beads \footnote{The following 
expression assumes that solids do not deform because of the capillary force. If solid deformation occurs, the numerical prefactor changes slightly
\cite{JKR_DMT}.} :
$F_{adh}\sim 2\pi \gamma R$, whatever the size of the (small) bridge
\cite{Israelashvili}.
Therefore, the adhesion force has no humidity dependence, and
would be moreover able to stick all beads together \cite{Duran}.
This apparent difficulty disappears if one takes into account
 the roughness of the beads.
At the liquid-vapour equilibrium, the  radius of curvature $r_{eq}$ 
of the liquid meniscus is fixed by the  so-called Kelvin relation  \cite{Laplace}
\begin{equation}
{\gamma \over r_{eq}} = \rho_l~k_BT~\log {P_{sat}\over P_v} \equiv
\rho_l~\Delta \mu
\label{Kelvin}
\end{equation}
where $\rho_l$ is the
density of the liquid and the ratio ${P_v\over P_{sat}}$ defines humidity. 
Under ambient conditions, this yields a nanometric order of magnitude
for $r_{eq}$ !  A crucial consequence is that liquid
bridges are able to form only in {\bf nanometric interstices}. Because the
beads are not smooth at the nanometer scale, the wetted region 
does not spread over the whole possible area it
would occupy if the beads were smooth,
 and the cohesive force is reduced in proportion.
Another counter-intuitive behaviour is the 
slow evolution in time of the
cohesive properties, measured through the time dependency of $\theta_w $.
 This indicates that condensation of liquid bridges
takes place over very long time scales. 
This feature is in agreement with experiments using Surface
Force Apparatus (SFA) and AFM techniques \cite{SFA}, which show
that an interstice between two solid surfaces can remain for
a very long time in a {\bf metastable} dry state 
while the equilibrium state would be a condensed liquid bridge.

On the basis of all these considerations, 
we propose a model based on the condensation of interstitial 
liquid bridges via an {\bf activated process}. 
In this letter, we shall only outline the main features of the physical
mechanism and a more detailed presentation 
 will be published elsewhere by the authors.
Let us consider two ``rough'' beads at contact. 
Under the assumption of an activated process, the time needed in order
to condense a liquid bridge in an interstitial volume is $\tau \simeq 
\tau_0~\exp\left({{\Delta E}\over 
{k_BT}} \right)$, with $\tau_0$ a microscopic time and $\Delta E$ an
activation energy barrier. Nucleation occurs preferentially at the level of
a nano-asperity, and the activation energy is accordingly $\Delta E \sim
\Delta \mu ~\rho_l a_0^2 e$, where  $\Delta \mu =\mu_{sat}-\mu_g\simeq
k_BT \log (P_{sat}/P_v)$, $\rho_l$ is the density of the liquid,
$e$ the gap between the surfaces at the level of the nucleating site
and $a_0^2$ a typical nucleation area (see fig. 1.b).
Now, since both beads are rough, 
one expects the nucleating sites to exhibit a broad statistics of gaps $e$ 
between solid surfaces and the activation times are accordingly widely
distributed. After a given time $t_w$, only the bridges with
an activation time $\tau_{act}$ smaller than $t_w$ have condensed. These
were therefore formed at the nucleating sites with a gap $e$ verifying 
$e<e_{max}(t_w)= {k_BT\over \Delta \mu} {1\over {\rho_l~a_0^2}}~\log \left(
{t_w \over \tau_0}\right)$.
Once a liquid bridge has condensed,  it
 fills locally the volume surrounding the nucleating site, 
until the Kelvin equilibrium condition for the radius of
curvature, eq. \pref{Kelvin}, is met. Thus, because of roughness, only a part of the total wettable area is indeed wetted at a given time $t_w$. 
The corresponding fraction $f(t_w)$, is proportionnal to the number of activated bridges, yielding in first approximation $f(t_w) \sim e_{max}(t_w)/\lambda$, 
with $\lambda$ the typical width of 
the distribution of distances between the surfaces. 
The capillary adhesion force is thus reduced by the same factor 
as compared to the perfectly smooth case, leading to  
\begin{equation}
F_{adh}(t_w)\simeq \gamma~d~{1\over \log {P_{sat}\over P_v}}~\log \left(t_w
\over \tau_0\right)
\label{Fadh}
\end{equation}
where $d=2\pi R/(\lambda \rho_l a_0^2)$ is a  distance taking into account the geometrical characteristics
of the contact.  Now, by reproducing Coulomb's argument for the stability of
the surface layer in the presence of this additional adhesive force, one
obtains the following implicit equation for $\theta_w(t_w)$ :
\begin{equation}
\tan \theta_w (t_w) \simeq \tan \theta_0~+~{\alpha (P_v)\over {\cos \theta_w (t_w)}} ~\log~\left({t_w\over t_0}\right)
\label{ageing}
\end{equation}
with $\alpha (P_v) =\alpha_0/\log {P_{sat}\over P_v}$ and $\alpha_0$ is a numerical constant depending on the characteristics of the
beads. Thus, by plotting $\tan \theta_w (t_w)$ as a function of 
$\log~(t_w)/\cos \theta_w (t_w)$, one should obtain a straight line :
as shown on fig. 2.a, 
this expectation is indeed in very good agreement with experimental results. 
Moreover,
as exhibited on fig. 2.b., the increase of the slope, $\alpha (P_v)$, of this line with
humidity is in good agreement with the theoretical prediction
eq. \pref{ageing}. 
Our model is thus able to reproduce both the waiting-time and humidity 
dependence of the measured ageing properties of a granular system.

Our results highlight the crucial role of humidity in the statics of 
granular systems. We have shown that the latter do exhibit ageing properties 
in an analogous way to
those encountered in dry friction for the static friction coefficient $\mu_s$.
The analogy is in fact
even more striking, since a similar effect of humidity has been reported
in friction between rocks \cite{Dieterich}, though not extensively
studied : in ref. \cite{Dieterich}, the ``standard'' ageing properties of $\mu_s$ were found
to disappear at vanishing humidity, just like in the case of a granular system !
Similar behavior were even observed in indentation experiments too 
\cite{indentation}.
It would be thus interesting to check to which extent the
analogy between dry friction and granular media is indeed pertinent.
The hope is to propose an alternative unsderstanding for the ageing properties
observed in solid friction.

Acknowledgements : The authors would like to thank Jean-Marie Georges
and M.L. Bocquet for highlighting discussions. L.B. thanks Jacques
Duran for kindly providing reference \cite{Coulomb}.

Correspondence and request for materials to L. Bocquet (e-mail : lbocquet@physique.ens-lyon.fr )

\newpage
{Figure 1.a : Description of the experimental setup. The glass beads fill $25~\%$ of the volume of a cylinder 
(diameter  $100 ~mm$, thickness $13~mm$) which can rotate
around its axes. Two different systems have been extensively studied, a ``polydisperse''
one, $140 \mu m <d< 260 \mu m $ and a ``monodisperse'' one $200 \mu m <d< 250 \mu m $, $d$ being the diameter of the beads. The walls of the cylinder are 
made of glass, and the shape
of the bead's pile is recorded with a video camera.
Experiments are performed at room temperature, 
under controlled humidity (defined as the ratio
between the vapour and saturated pressure of water, $P_v/P_{sat}$). 
Before starting experiments, the system is prepared by rotating the
drum during approximatively 12 hours. Then the ageing properties
are investigated for various values of $P_v/P_{sat}$ using the
following  protocol : 
(i) first, the system is put in motion
for a few turns (typically three); (ii) the end of this ``motion period''
defines the origin of waiting time $t_w=0$, after which the
system is left at rest; (iii) after a given time (ranging from
$10$ to $10^4$ seconds), a slow rotational motion (with velocity less than 
one turn per minute) is transmitted to the
cylindrical drum. The angle $\theta_w$ at which the first avalanche takes
place is measured from the slope of the beads
just before the avalanche. The same procedure is repeated for different waiting
times $t_w$ and a whole curve $\theta_w$ {\it vs.} waiting time $t_w$ can
be constructed.
\label{fig1}
}

{Figure 1.b : Schematic drawing of the contact at the nanometer scale
between two micrometric asperities on the beads. Inside most of the contact region,
the solid surfaces do not really touch eachother at the molecular scale,
and capillary condensation occurs in the interstice left between the surfaces.
\label{fig2}
}

{Figure 2.a : Logarithmic ageing of the angle of first avalanche. 
As explained in the text after eq. \pref{ageing}, the ageing property is best
analysed by plotting $\tan \theta_w (t_w)$ as a function of $\log_{10} (t_w)/\cos
\theta_w (t_w)$, for different values of the water
vapor pressure $P_v$.
From bottom to top, humidity is : $15 \%$ (triangles), $27 \%$
(pentagons), $36.1\%$ (squares) and
$45.5 \%$ (circles). In these measurements, times runs typically over the
range $t=5~s$ up to $t=5000~s$. The dotted lines are least-square fits
of the experimental data, whose slope is identified with $\alpha(P_v)$.

\label{fig3}
}

{Figure 2.b : Variation of the slope $\alpha(P_v)$ characterizing the
ageing behavior of the first avalanche angle (see text eq. \pref{ageing})
with humidity $P_v/P_{sat}$. Open dots correspond to the polydisperse
system, and filled dots to the monodisperse one. The dashed line is the theoretical 
prediction $\alpha=\alpha_0/\log(P_{sat}^{*}/P_v)$, where  
$\alpha_0=0.079$ and $P_{sat}^{*}=0.68~P_{sat}$. The validity of the previous
theoretical prediction is best confirmed by the measured linear
dependence of $1/\alpha$ as a function of
$\log(P_{sat}/P_v)$ (see inset) : in this plot, a linear least-square fit thus provides
unambiguously both values of $\alpha_0$ and $P_{sat}^{*}$.
The lowering observed in the saturating pressure $P_{sat}^{*}$ might be an
effect of the long-ranged attraction forces exerted by the walls,
and/or of dissoluted species in condensed water.  
\label{fig4}
}

\newpage
\begin{figure}
$$\input{viell_2.pstex_t}$$
\vskip 2cm
Figure 1a. Moisture induced Ageing in Granular Media

\noindent Corresponding author : L. Bocquet 
\end{figure}

\begin{figure}
$$\input{viell_1.pstex_t}$$
\vskip 2cm
Figure 1b. Moisture induced Ageing in Granular Media

\noindent Corresponding author : L. Bocquet 
\end{figure}

\newpage
\begin{figure}
\begin{center}
\setlength{\unitlength}{1mm}
\begin{picture}(100,100)
\epsfig{file=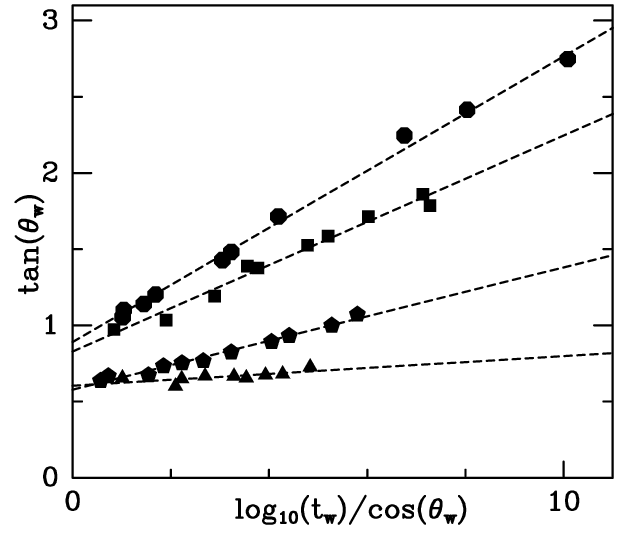}%
\end{picture}
\end{center}
\vskip 2cm
Figure 2a. Moisture induced Ageing in Granular Media

\noindent Corresponding author : L. Bocquet 
\end{figure}

\newpage
\begin{figure}
\begin{center}
\setlength{\unitlength}{1mm}
\begin{picture}(100,100)
\epsfig{file=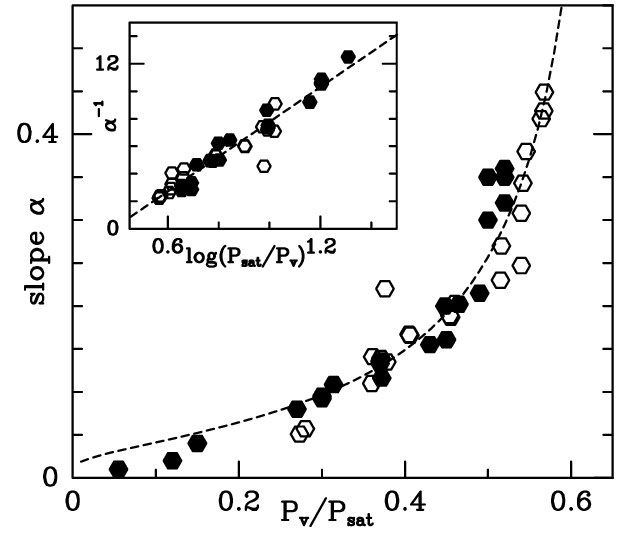}%
\end{picture}
\end{center}
\vskip 2cm
Figure 2b. Moisture induced Ageing in Granular Media

\noindent Corresponding author : L. Bocquet 
\end{figure}

\end{document}

%% file: viell_2.pstex_t
\begin{picture}(0,0)%
\epsfig{file=viell_2.pstex}%
\end{picture}%
\setlength{\unitlength}{0.00083300in}%
\begingroup\makeatletter\ifx\SetFigFont\undefined
\def\x#1#2#3#4#5#6#7\relax{\def\x{#1#2#3#4#5#6}}%
\expandafter\x\fmtname xxxxxx\relax \def\y{splain}%
\ifx\x\y   
\gdef\SetFigFont#1#2#3{%
  \ifnum #1<17\tiny\else \ifnum #1<20\small\else
  \ifnum #1<24\normalsize\else \ifnum #1<29\large\else
  \ifnum #1<34\Large\else \ifnum #1<41\LARGE\else
     \huge\fi\fi\fi\fi\fi\fi
  \csname #3\endcsname}%
\else
\gdef\SetFigFont#1#2#3{\begingroup
  \count@#1\relax \ifnum 25<\count@\count@25\fi
  \def\x{\endgroup\@setsize\SetFigFont{#2pt}}%
  \expandafter\x
    \csname \romannumeral\the\count@ pt\expandafter\endcsname
    \csname @\romannumeral\the\count@ pt\endcsname
  \csname #3\endcsname}%
\fi
\fi\endgroup
\begin{picture}(6832,5668)(2611,-7710)
\end{picture}

%% file: viell_1.pstex_t
\begin{picture}(0,0)%
\epsfig{file=viell_1.pstex}%
\end{picture}%
\setlength{\unitlength}{0.00083300in}%
\begingroup\makeatletter\ifx\SetFigFont\undefined
\def\x#1#2#3#4#5#6#7\relax{\def\x{#1#2#3#4#5#6}}%
\expandafter\x\fmtname xxxxxx\relax \def\y{splain}%
\ifx\x\y   
\gdef\SetFigFont#1#2#3{%
  \ifnum #1<17\tiny\else \ifnum #1<20\small\else
  \ifnum #1<24\normalsize\else \ifnum #1<29\large\else
  \ifnum #1<34\Large\else \ifnum #1<41\LARGE\else
     \huge\fi\fi\fi\fi\fi\fi
  \csname #3\endcsname}%
\else
\gdef\SetFigFont#1#2#3{\begingroup
  \count@#1\relax \ifnum 25<\count@\count@25\fi
  \def\x{\endgroup\@setsize\SetFigFont{#2pt}}%
  \expandafter\x
    \csname \romannumeral\the\count@ pt\expandafter\endcsname
    \csname @\romannumeral\the\count@ pt\endcsname
  \csname #3\endcsname}%
\fi
\fi\endgroup
\begin{picture}(6394,6482)(2188,-6902)
\end{picture}